# Unveiling the secrets of the mid-infrared Moon

Yunzhao Wu[1,2], Qi Jin[2], Cui Li[1,3], Tianyi Xu[2], Wenwen Qi[4], Wei Tan[4], Xiaoman Li[4], Zhicheng Shi[4], Hongyan He[4], Shuwu Dai[5], Guo Li[5], Fengjing Liu[5], Jingqiao Wang[6], Xiaoyan Wang[6], Yu Lu[1], Wei Cai[1,7], Qi Wang[8], Lingjie Meng[8]

[1]Key Laboratory of Planetary Sciences, Purple Mountain Observatory, Chinese Academy of Sciences, Nanjing, 210034, China; wu@pmo.ac.cn

[2]Space Science Institute, Macau University of Science and Technology, Macau, China

[3]CAS Center for Excellence in Comparative Planetology, China

[4]Beijing Institute of Space Mechanics and Electricity, Beijing, China

[5]Beijing Institute of Spacecraft System Engineering, Beijing, China

[6]China Centre For Resources Satellite Data and Application, Beijing, China

[7]Jiangsu Center for Collaborative Innovation in Geographical Information Resource Development and Application, Nanjing, 210023, China

[8]Earth Observation System and Data Center, China National Space Administration, Beijing, China

**Abstract**

The Moon's optical characteristics in visible and long-wavelength infrared (LWIR) have long been observed with our eyes or with instruments. What the mid-infrared (MIR) Moon looks like is still a mystery. For the first time we present detailed appearance of the MIR Moon observed by a high-resolution geostationary satellite and reveal the essence behind its appearance. The appearance of the MIR Moon is opposite to its normal visible appearance. In addition the MIR Moon shows limb darkening. Both the absolute and the relative brightness distribution of the MIR lunar disk changes with the solar incidence angle. The signatures of the MIR Moon are controlled by both the reflection and emission of the lunar surface. We also show first-ever brightness temperature maps of the lunar disk without needing a mosaic, which better show the temperature variation across the lunar disk. They reveal that the relationship between brightness temperature and solar incidence angle $i$ is $\cos^{1/b}i$, and the power parameter is smaller than the Lambertian temperature model of $\cos^{1/4}i$ observed for lunar orbit-based measurements. The slower decrease of the brightness temperature when moving away from the sub-solar point than the Lambertian model is due to topographic effects. The brightness temperature is dominated by albedo and the solar incidence angle and influenced by the topography. Our results indicate that the Moon in the MIR exhibits many interesting phenomena which were previously unknown, and contains abundant information about lunar reflection and thermal emission for future study.

*Unified Astronomy Thesaurus concepts:* The moon (1692); Geostationary satellites (647); Lunar phase (968); Infrared astronomy(786)

# 1. Introduction

The Moon's optical characteristics are its most accessible properties, which have been long studied. Our eyes sense the Moon at wavelengths in the visible spectrum (0.38-0.76 μm). By using instruments the knowledge of the Moon's spectral properties has expanded to include the near-infrared (NIR: 0.76-1.4 μm) (Buratti et al. 1996; Robinson & Riner 2005; Ohtake et al. 2010), short-wavelength infrared (SWIR: 1.4-3.0 μm) (Kouyama et al. 2016; Shkuratov et al. 2011; McCord et al. 1981; Kieffer & Stone 2005; Stone et al. 2002; Pieters et al. 2009; Velikodsky et al. 2011; Wu & Hapke 2018) and the long-wavelength infrared (LWIR: 5-1000 μm) (Pettit & Nicholson 1930; Saari et al. 1972; Lawson et al. 2000; Paige et al. 2010; Nash et al. 2017; Vasavada et al. 2012; Williams et al. 2017; Shirley & Glotch 2019; Hanna et al. 2017; Hayne et al. 2017). Despite some research at low spatial resolution (Clark 2009; Sunshine et al. 2009), many of the Moon's characteristics in the mid-infrared (MIR: 3-5 μm) remain to be discovered. What does the Moon look like at MIR wavelengths? How does it change with the lunar phase? What is the essence behind its appearance? Observations with higher spatial resolution are necessary to reveal the Moon's properties in this range of wavelengths.

In this paper, we show the MIR Moon's appearance and its variation with lunar phase observed by China's high-resolution geostationary satellite, Gaofen-4 (GF-4), and reveal the essence behind its appearance. The paper is organized as follows: in Section 2, we introduce the observation and data; in Section 3 we present our results, including the appearance of the MIR Moon and the first-ever detailed temperature map of the lunar disk; Section 4 is a discussion of our findings, including the separation of the reflected and emitted radiations, and distribution characteristic of temperature over the lunar disk; Section 5 is our summary.

# 2. Data

At 00:04 Beijing time on December 29, 2015, GF-4 was launched from the Xichang Satellite Launch Centre using a Long March 3B rocket. GF-4 is a geostationary satellite, part of the China High-resolution Earth Observation System (CHEOS). The task of GF-4 is to offer visible and near-infrared (VNIR) spatial resolution of better than 50 m and mid-infrared (MIR) spatial resolution of 400 m from geostationary orbit. GF-4 has a large-array VNIR detector and an MIR detector with fields of view (FOV) of 0.8° × 0.8° and 0.66° × 0.66°, respectively, and instantaneous fields of view (IFOV) of 1.363 and of 11.249 μrad/pix, respectively. Based on the large-array detectors, the entire lunar disk was imaged in both VNIR and MIR in a single exposure with spatial resolutions of up to ~500 m/pixel and ~4 km/pixel, respectively. Such high-resolution single-exposure images are advantageous for investigating the brightness/temperature distribution across the lunar disk compared to a mosaic of image frames or scan-line imagery, such as the LWIR camera onboard Clementine (Lawson et al. 2000) and the Diviner Radiometer onboard the Lunar Reconnaissance Orbiter (LRO) (Paige et al. 2010; Nash et al. 2017; Vasavada et al. 2012; Williams et al. 2017).

    GF-4 imaged the Moon in the MIR with the best spatial resolution achieved to date on the day of China's Lantern Festival (March 2, 2018). Then before and after the day of the century's longest lunar eclipse (July 28, 2018), the second batch of lunar MIR images were taken with phase angles of approximately ±30°, 90° (half Moon), and 0° (full Moon). For each imaging of the Moon we observed in the VNIR (bands 1–5) and the MIR (band 6). The spectral range of band 6 is 3.50–4.10 μm with the lunar effective wavelength of 3.77 μm. This wavelength marks the turning point at which the reflectance of lunar soil starts to decrease: according to laboratory spectra of lunar soils, the wavelength at which lunar soil has the largest reflectance is around 3.8–4.0 μm. Hence, the GF-4 provides an excellent opportunity to show in detail the combined effects of reflected sunlight and thermal emission of the Moon. Tables 1 and 2 provide details about the instrument performances and the measurement geometry. Appendix A provide details about the calibration and temperature calculation.

### 3. Results

*3.1. The MIR Moon's appearance and variation with lunar phase*

    Figure 1 shows the Moon in the NIR and MIR range imaged by GF-4 over 3 days. The images of March 2 and August 4 are not shown because they were uncalibrated. The NIR lunar images are similar to visible images. (The images of all the 5 days are

shown in Appendix B). At first glance the Moon in MIR appears inverted compared to its image in VNIR. On the MIR lunar disk, the low albedo surfaces, such as maria, become brighter than the highlands, and the high albedo features, such as fresh craters, crater rays, and the Reiner Gamma swirl, become darker than their surroundings. The Moon in the VNIR shows a relatively uniform brightness (Lommel-Seeliger behavior) while the Moon in the MIR exhibits limb darkening across the disk. The Moon in the MIR shows a diminished contrast between maria and highlands, and fresh and mature materials are not as sharp as those in the VNIR Moon. Crater rays appear dimmer than those in the VNIR Moon. In contrast, topographies such as craters, domes, wrinkle ridges, and Mons are obvious even near zero phase angle. These features indicate that at MIR wavelengths both the reflected sunlight and the thermal emission from the Moon itself are significant. The higher the albedo of the surface the lower its thermal emission: thus, the combination of these two opposing effects causes the reduced contrast in the MIR.

The brightness distribution of the Moon in the MIR changes with the local time. The brightest location follows the Sun and is not exactly at the sub-solar point but in the nearby mare. The visibility of crater rays also varies with the solar angle. The closer to the sub-solar point they are, the clearer they become. The clearest crater ray system in the VNIR, Tycho, is almost invisible in the MIR because it is far from the equator. The obvious crater ray systems in the MIR are those at low latitudes, and the two most obvious craters are Copernicus and Kepler. As shown in the MIR images of July 25 and August 4, when the sub-solar point is far from the two craters, the crater rays are invisible. This indicates that in the MIR, the proportion of thermal emission decreases and the proportion of surface reflection increases as one moves away from the sub-solar point. Mare Tranquillitatis has a lower albedo than Mare Serenitatis. Hence, Mare Tranquillitatis is brighter than Mare Serenitatis even for a solar incidence angle $i = 45°$ (MIR image of July 28). Mare Tranquillitatis becomes darker than Mare Serenitatis when the solar incidence angle is $i = 60°$ (MIR image of July 30). This means that for the Moon in the MIR both the absolute brightness distribution and the relative brightness among regions vary with local time within one lunar day, demonstrating the effect of the solar insolation.

### 3.2. Temperature map of the whole lunar disk

Figure 2 shows the brightness temperature maps. As with the image in the MIR, the brightness temperature image also clearly shows the topography. The brightness temperature is approximately concentrically distributed around the sub-solar point. For identical solar angles the low-albedo region has a higher temperature than the high-albedo region. The brightness temperature differences are the results of the albedo variations rather than thermal inertia. The location of the highest temperature is not exactly at the sub-solar point but in the nearby mare, because in the examples we show, the sub-solar point is always located at a short distance from a mare. The highest brightness temperatures of the three days are 383.99, 386.35 and 385.52 K with radiometric calibration uncertainty of ±2.25% (1σ standard deviation, Appendix A). The highest temperature measured by channel 7 of Diviner is ~390 K (Williams et al. 2017) and the maximum bolometric brightness temperatures measured by Diviner are ~387–397 K with an average of 392.3 K (Shirley & Glotch 2019). The region that has the highest temperature is almost the same as that of the highest radiance of the Moon in the MIR. This indicates that around noon the thermal emission dominates the lunar brightness in the MIR. The basalts with the lowest albedo have a higher temperature than most highlands, even at a 45° solar incidence angle. This indicates that all three types of lunar disk images (VNIR, MIR, and brightness temperature) are controlled by albedo.

Figure 3 shows the histogram of all three types of lunar image. The absolute brightness of Moon in the NIR is much higher than in the MIR. At full Moon the highest brightness in the NIR can be as high as 90 W/m$^2$/sr/μm while in the MIR it is less than 9 W/m$^2$/sr/μm. The brightness in the NIR depends on the lunar phase while the brightness in the MIR was relatively stable over the three days with the highest brightness <9 W/m$^2$/sr/μm. For the brightness temperature, all phases exhibit a negatively skewed distribution and the highest values are also similar. At full Moon the VNIR band exhibits a bimodal distribution which corresponds to the maria and highlands, while the MIR band exhibits an approximately uniform distribution, reflecting the diminished contrast between maria and highlands. The histogram of the –30° phase (waxing phase of July 25) in the VNIR is similar to that of full Moon, while that of 30° phase (waning phase of July 30) is quite different with a positively skewed distribution. This reflects the fact that the lunar western hemisphere has more maria than its eastern hemisphere.

## 4. Discussion

*4.1. Separation of the reflected and emitted radiations*

The fundamental vibration bands of some minerals and life-related molecules such as water and organic matter occur at MIR wavelengths, so that the separation of the reflected sunlight from the thermal emission is conducive to the research of these materials. The simultaneous imaging of all six bands allowed for a direct comparison between the Moon in the VNIR and MIR and the separation of the reflection and thermal emission from the Moon in the MIR. Figure 4 shows the contribution of the reflective and emissive lunar radiance. At 3.77 μm for most of the Moon the thermal emission is greater than the reflected solar radiance. The emitted fraction of the maximum temperature area over the three days is 84.7±0.25%. The thermal emission varies considerably from 0 to 8 W/m$^2$/sr/μm, whereas the reflected fraction usually varies between 1 and 2 W/m$^2$/sr/μm, which causes both types of radiance to be visually unrelated. However, the right side of the scatter plot shows a fan-like shape with a pointed tip on the extreme right: the high-albedo side of the envelope indicates a negative correlation, showing that a larger thermal emitted radiance corresponds to a smaller reflected radiance.

*4.2. Distribution of temperature over the disk*

Since the early 20th century, many efforts have been made to investigate the temperature variation across the lunar disk. Pettit and Nicholson (1930) found that the center-to-limb temperature variation as a function of the local solar incidence angle *i* was $\cos^{1/6} i$ rather than the $\cos^{1/4} i$ variation expected for a Lambertian surface. Based on 23 scans of the sunlit portion of the surface through a lunation, Saari and Shorthill (1972) suggested a replacement of the $\cos^{1/6} i$ law by an expression which is linear in cos*i*.

Conversely, using the Clementine LWIR camera images, Lawson et al. (2000) found that the Lambertian temperature model is a fair approximation for nadir-looking temperatures, which is supported by Diviner (Williams et al. 2017). Note that none of previous studies imaged the entire lunar disk with a single exposure. The high-resolution multiband images of GF-4 taken as single exposures provide a good opportunity to illuminate in detail the center-to-limb variability for both reflection and temperature. Figure 5 shows the radiance and brightness temperature as a function of local solar incidence angle. To avoid the effects of compositional variation, the profiles were carefully selected from compositionally homogeneous regions for both highlands and maria. The profile across the lunar surface of the Moon in the MIR (middle panel) is smoother than that of the Moon in the VNIR (top panel), which is consistent with the diminished contrast of the Moon in MIR. The larger minimum-maximum variation in the highlands in both VNIR and MIR bands compared to maria indicates a greater roughness of the highlands. The profiles confirms relatively uniform brightness across the lunar disk for the VNIR range and limb darkening for both the MIR range and brightness temperature. But neither of the two types of lunar limb darkening obeys the Lambertian model. For the MIR band, the relationship between radiance and solar incidence angle is linear with a slope of -0.07 W/m$^2$/sr/µm/deg for the highlands and -0.1 W/m$^2$/sr/µm/deg for the maria. For brightness temperature, the local temperature T of highlands varies with solar incidence angle $i$ as:

$$T=T_{ss}\cos^b i \qquad (1)$$

where $T_{SS}$ is the temperature of the sub-solar point and b is a parameter to be fit. For a Lambert sphere b is 1/4. The GF-4 satellite reveals that for highlands $T_{SS}$ and b are 372 and 1/6 for July 25 and 373 and 1/7 for July 28, and for maria $T_{SS}$ and b are 377 and 1/6 for July 25 and 373 and 1/12 for July 28. Note that the profile for maria for full Moon varies more linearly as T=319.72+58.62 cos$i$. One of the reasons might be the difficulty in selecting compositionally homogeneous mare regions. Anyway, none of these profiles obey the Lambertian model. This indicates that the decrease of the brightness temperature when moving away from the sub-solar point is slower than would be described using a Lambertian model. This is due to topographic effects: at large solar incidence angle the sunward facing slopes enhanced thermal emission compared to a flat surface. The topographic effects become stronger as the solar incidence angle increases. Figure 5 shows that at solar incidence angle >60º at full Moon the average topography should be tilted toward the Sun by angles of ~10º relative to a horizontal surface.

Lawson et al. (2001) found that as the phase angle increases the influence of surface roughness grow. However, the emission angle of Clementine is 0˚ and the phase angle is equal to the solar incidence angle. That is, the influence of the surface roughness as a function of phase angle derived from Clementine data is actually the solar incidence angle. The large variations of both solar incidence angle and emission angle across the whole lunar disk derived from GF-4 observations better illuminate the effects of surface roughness as a function of various geometries. Figure 5 shows that for both maria and highlands the temperature profile as a function of solar incidence angle at full Moon (small lunar phase) deviates more from the Lambertian model than that at large lunar phase (July 25), suggesting that the influence of surface roughness is stronger at small lunar phase rather than at large lunar phase. This is understandable. Compared to large lunar phase, observations at full Moon have less shadow within the field of view and hence more enhanced thermal emission.

In summary, solar incidence angle dominates the thermal emission, but phase angle also has an influence. We further selected eight homogenous areas that were observed multiple times to strengthen the understanding of their variations as a function of observation geometry. These areas, two sub-camera points, three sub-solar points, and three calibration sites (Apollo 16, Mare Serenitatis (MS2) (Pieters et al. 2008) and Chang'E-3 (CE-3) (Wu et al. 2018)), can be considered representative of the whole Moon because they cover highlands and maria, low and high latitude, sub-camera and sub-solar points. As shown in Figure 6, the brightness temperature reduces with increasing solar incidence angle, and when the solar incidence angle is <40º the reduction is fast - closer to a Lambertian temperature model, and when the solar incidence angle is large the reduction is relatively slow. It confirms that at large incidence angle the topographic effect becomes more significant. The profiles of MS2, sub-solar point of July 25 and sub-solar point of July 30 show that although the full Moon almost has a phase angle closer to $0°$, the highest temperature still occurs at the smallest solar incidence angle. However, the temperature reduction of the full Moon is relatively small, which indicates the influence of phase angle. This is especially obvious when the difference of solar incidence is small. For example, for both Apollo 16 and CE-3 the highest temperatures occur at full Moon rather than at the smallest solar incidence angle. In the VNIR bands the phase angle plays a dominant role. For all the eight areas, the largest values always occur at full Moon and are much larger than at other phases, which represents an opposition surge (Appendix C). For Apollo 16 and CE-3 the difference of the incidence angle between full Moon and non-full Moon is only $1°$ but the radiance of the full Moon is 1.6-1.7 times that of the non-full Moon for Apollo 16 and CE-3, respectively.

## 5. Summary

The Moon in the MIR exhibits many interesting phenomena which were previously unknown. It contains abundant information about lunar reflection and thermal emission. The GF-4 high-resolution observations of the entire lunar disk in a single exposure offer a unique thermal emission perspective of the lunar hemisphere, and the data illuminate global temperature variations with local incidence and emission angles that are not available to lunar-orbiting spacecrafts and not restricted by the terrestrial atmosphere. When compared to the previous models on thermal spatial variability, the high-resolution MIR lunar images together with the simultaneously imaged VNIR Moon directly illustrate it. Both the reflectance and emissivity are strongly dependent on the illumination and observation geometry. The high-spatial resolution observations of the whole lunar disk provide an unprecedented opportunity for future work building a directional distribution model for both reflectance and emissivity. Moreover, more knowledge on lunar reflection and thermal mission, such as surface roughness and thermal inertia, can be derived by future long-term data analysis.

**Acknowledgments:**

We would like to thank China National Space Administration (CNSA) for organizing the observations and Greg Michael for English editing. This research was supported by the National Key R&D Program of China (2018YFB0504700), the Macau Science and Technology Development Fund (103/2017/A, 119/2017/A3) and Minor Planet Foundation of Purple Mountain Observatory.

**Table 1**
The instrument performance parameters of GF-4

| Bands | Wavelength range/μm | Pixel number | Lunar effective wavelength/μm |
|---|---|---|---|
| Band 1 – Panchromatic | 0.45–0.90 | 10 k × 10 k | 0.61 |
| Band 2 – Blue | 0.45–0.52 | 10 k × 10 k | 0.49 |
| Band 3 – Green | 0.52–0.60 | 10 k × 10 k | 0.56 |
| Band 4 – Red | 0.63–0.69 | 10 k × 10 k | 0.65 |
| Band 5 – NIR | 0.76–0.90 | 10 k × 10 k | 0.81 |
| Band 6 – MIR | 3.50–4.10 | 1 k × 1 k | 3.77 |

**Table 2**
The geometry of the 5 observations

| Time (UTC) | Moon–Sun distance (Au) | Moon–Camera distance ($10^4$ km) | Subsolar (°) | | Subcamera (°) | | Phase angle (°) |
| | | | Lat | Lon | Lat | Lon | |
|---|---|---|---|---|---|---|---|
| 2018.03.02T04:00:25 | 0.991 | 40.798 | -0.75 | 2.16 | -1.68 | 2.37 | -0.95 |
| 2018.07.25T02:49:02 | 1.016 | 44.437 | -0.11 | 32.83 | -5.68 | 3.21 | -30.09 |
| 2018.07.28T04:49:00 | 1.015 | 44.581 | -0.02 | -4.77 | -1.47 | -1.17 | 3.88 |
| 2018.07.30T06:49:20 | 1.015 | 44.439 | 0.03 | -30.18 | 2.08 | -3.33 | 26.92 |
| 2018.08.04T12:25:20 | 1.015 | 42.063 | 0.18 | -94.06 | 8.42 | -5.00 | 89.04 |

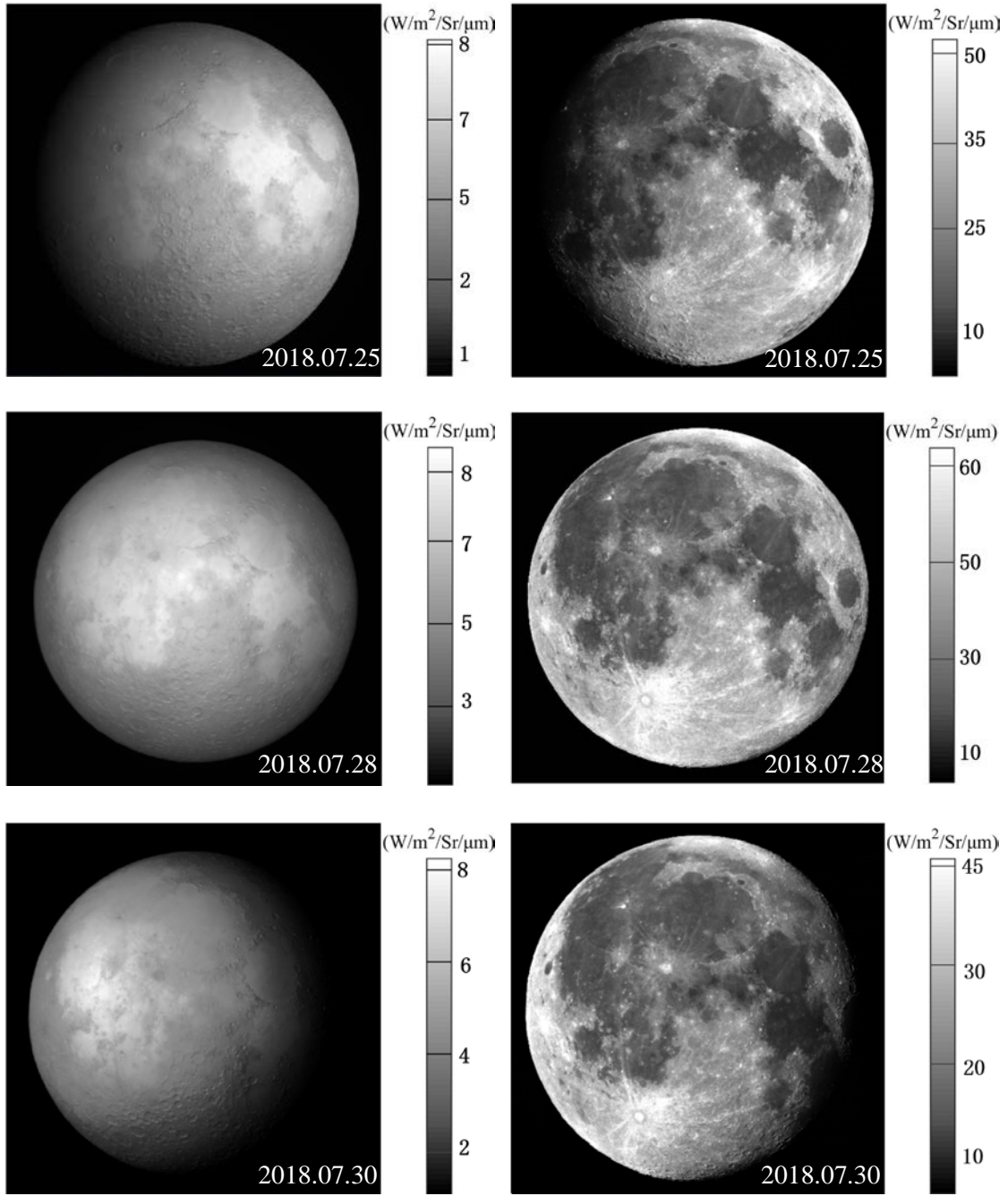

**Figure 1.** The radiance of the Moon imaged by the GF-4 band 6 (left column) and band 5 (right column) for 3 days.

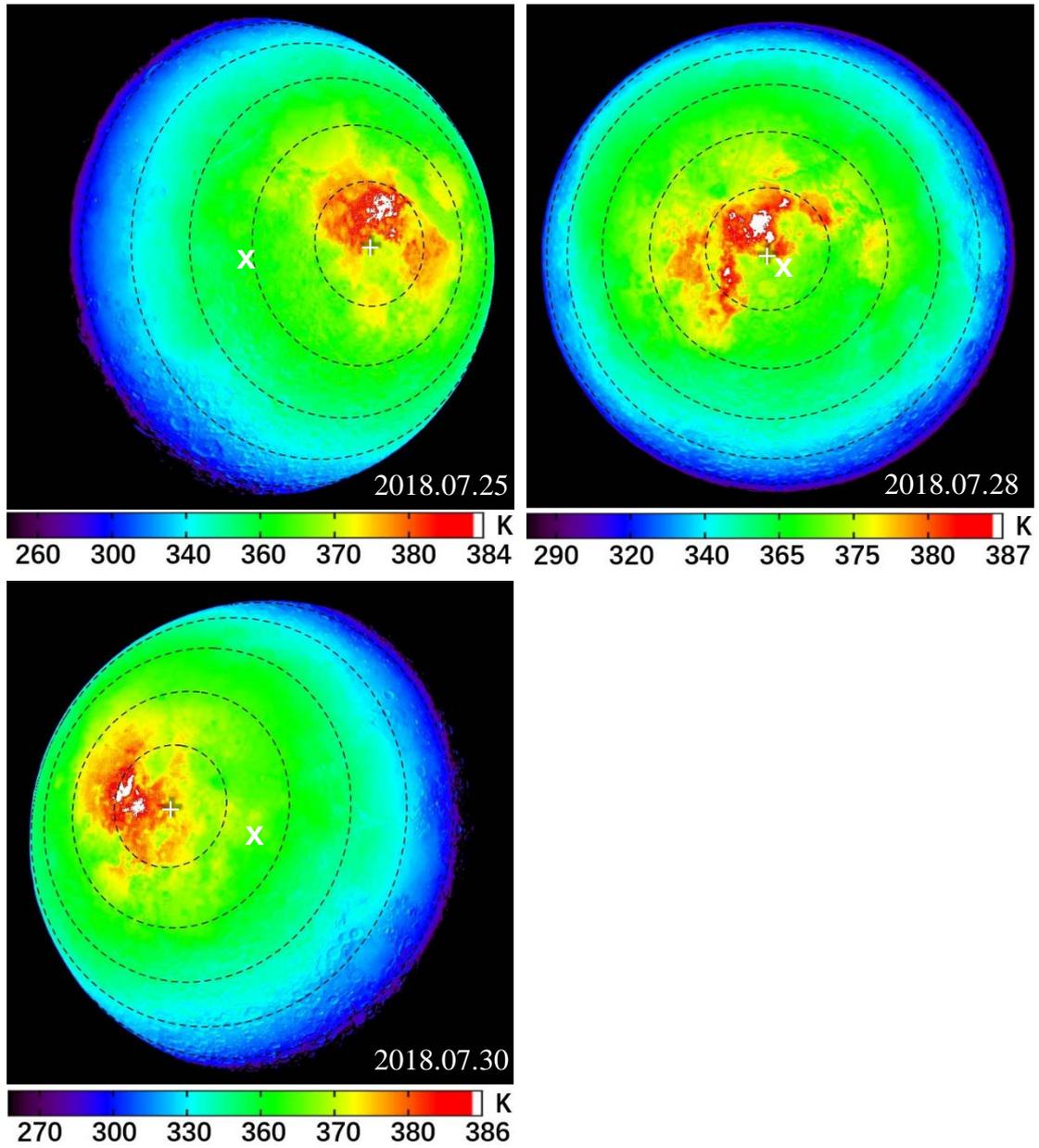

**Figure 2.** Brightness temperature distribution of the Moon disk. "+" represent sub-solar and "×" represent sub-camera points, respectively. Dashed lines represent solar incidence angles with an interval of 15°.

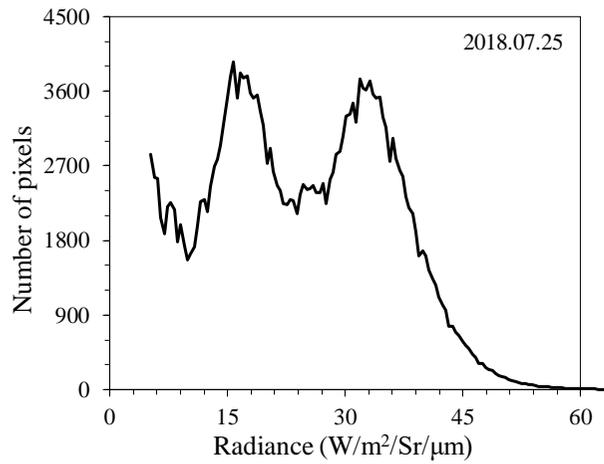 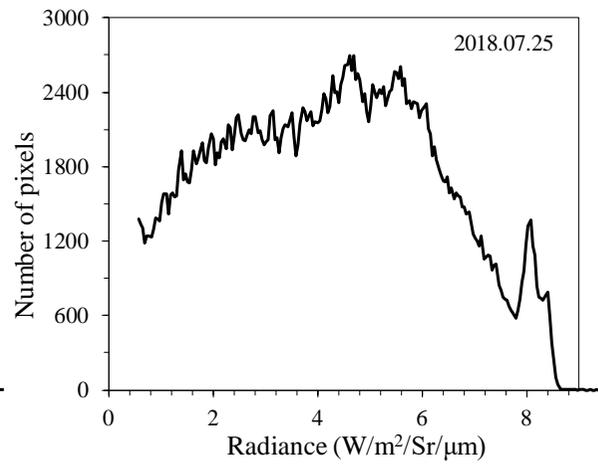 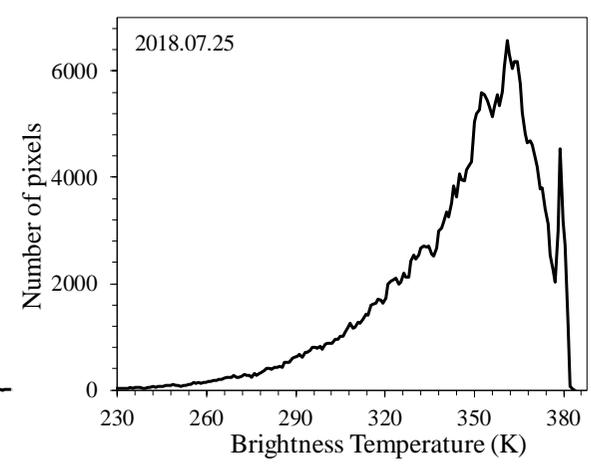
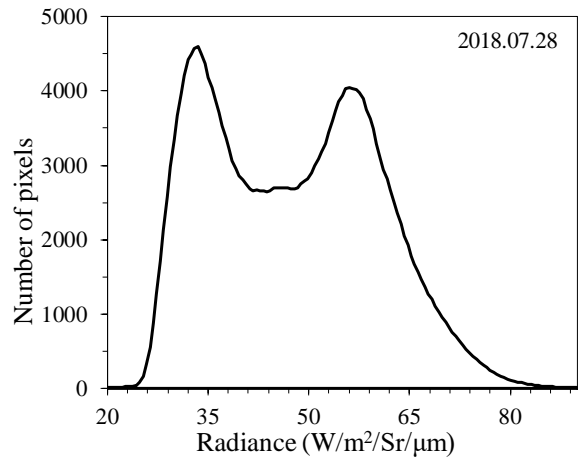 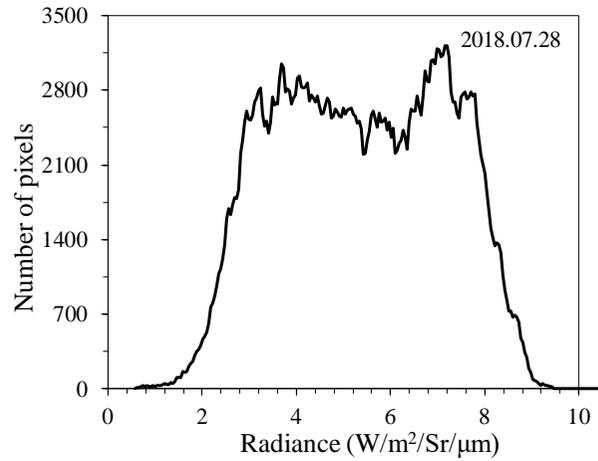 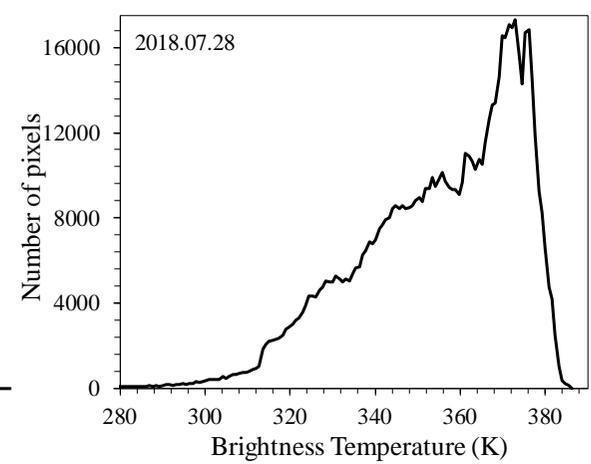

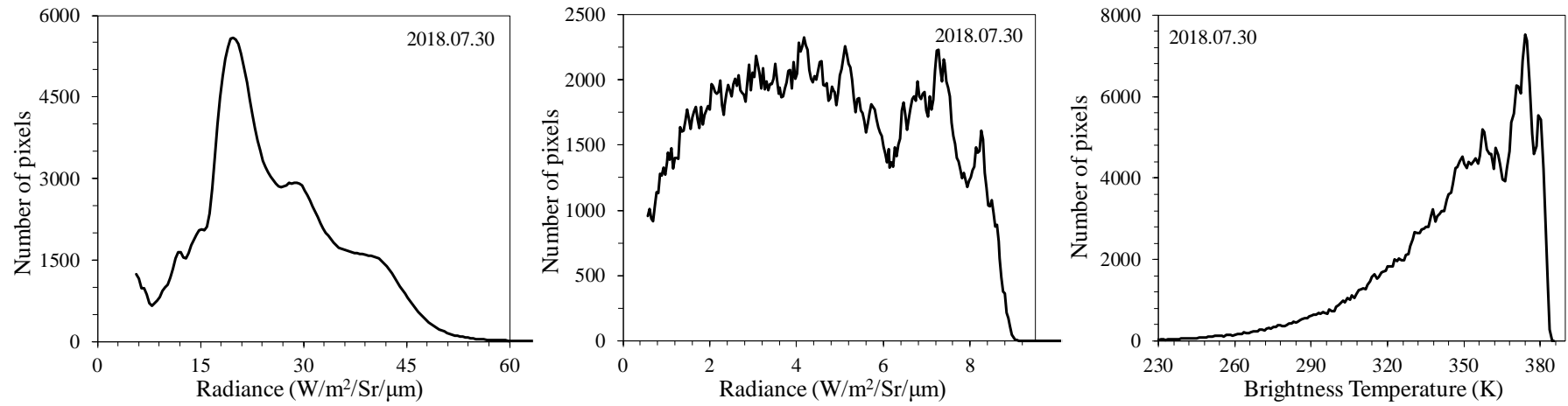

**Figure 3.** Histograms of VNIR, MIR, and brightness temperature of the lunar disks.

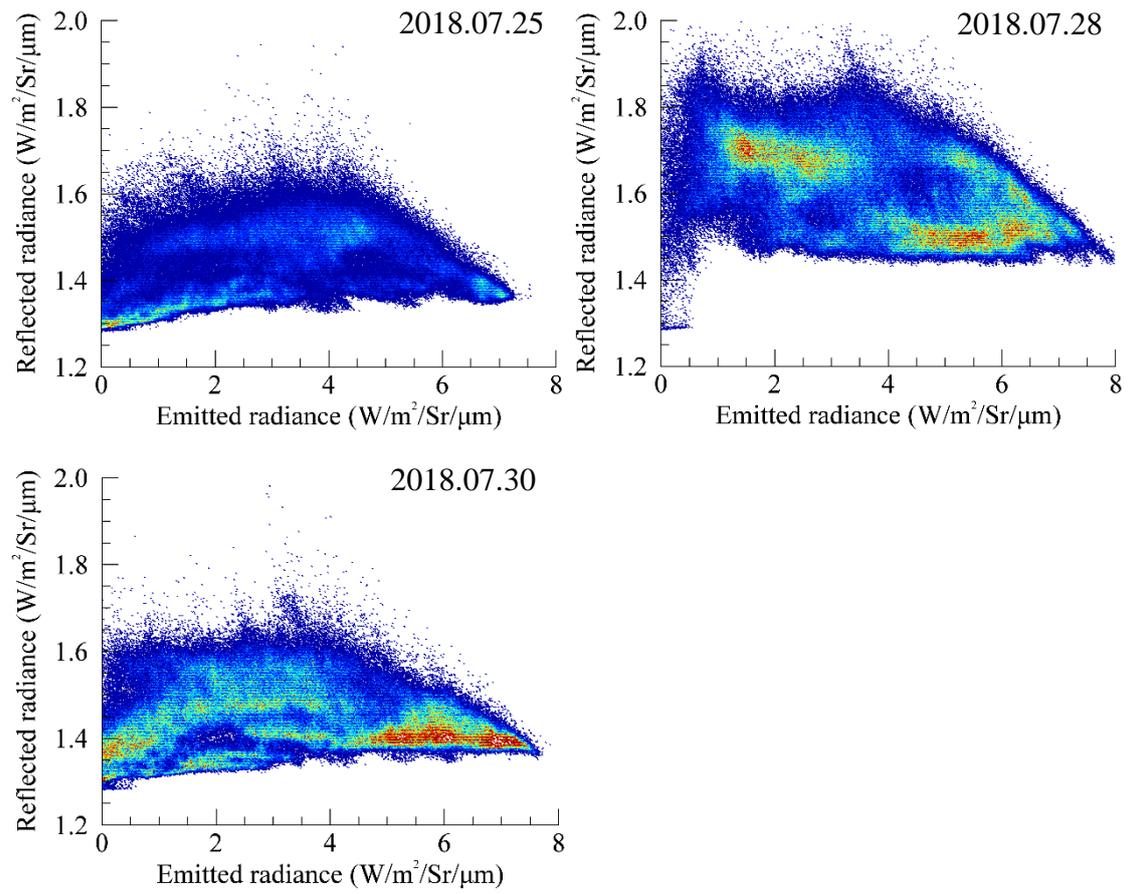

**Figure 4.** Reflective and emissive lunar radiance.

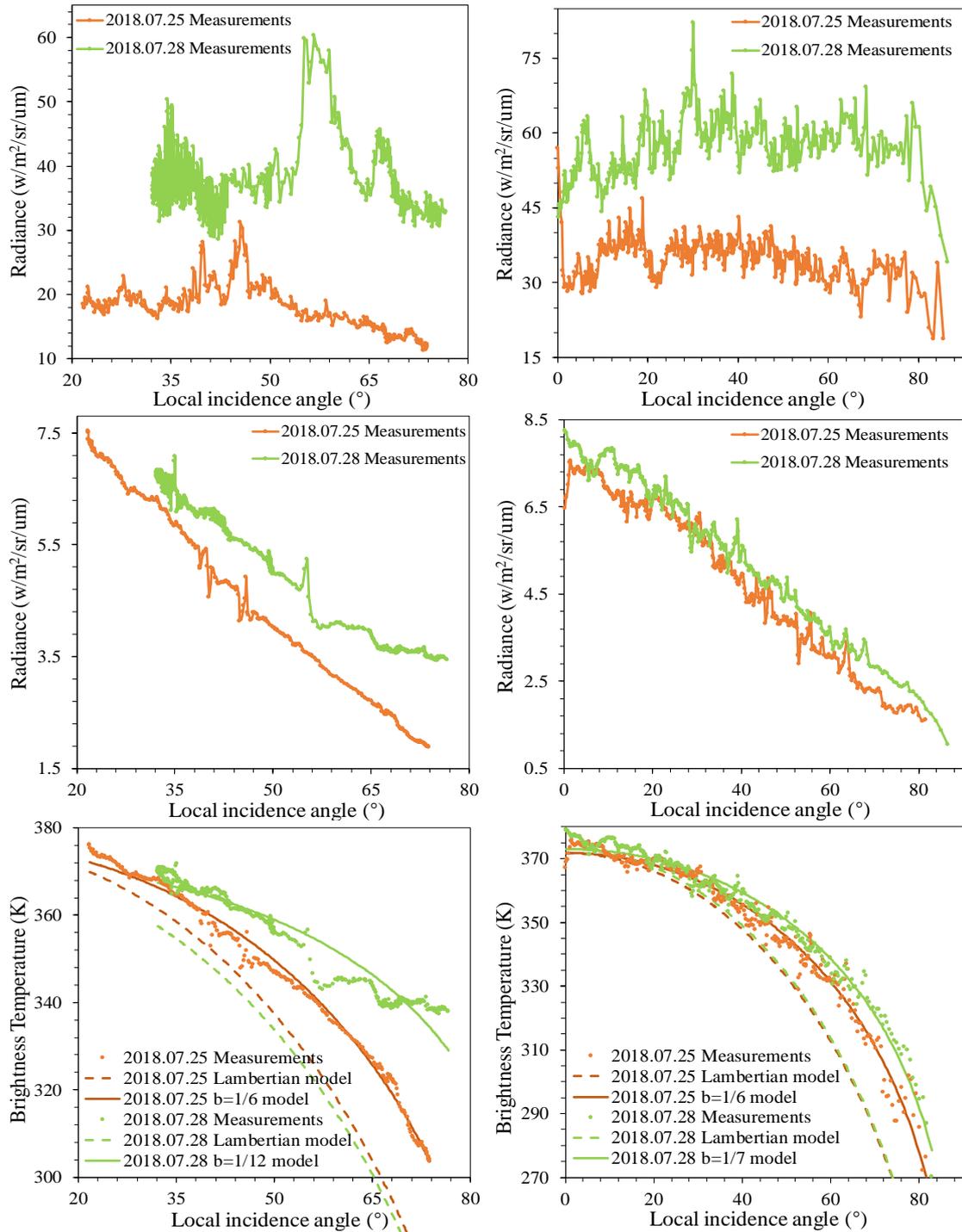

**Figure 5.** Radiance and brightness temperature versus local incidence angle. Left column: Maria. Right column: Highlands. Top: band 5. Middle: band 6. Bottom: brightness temperature.

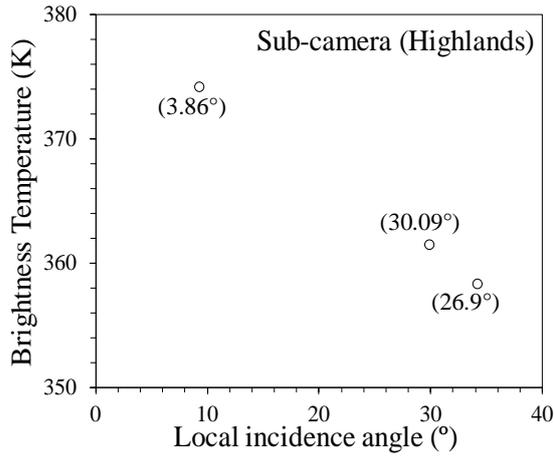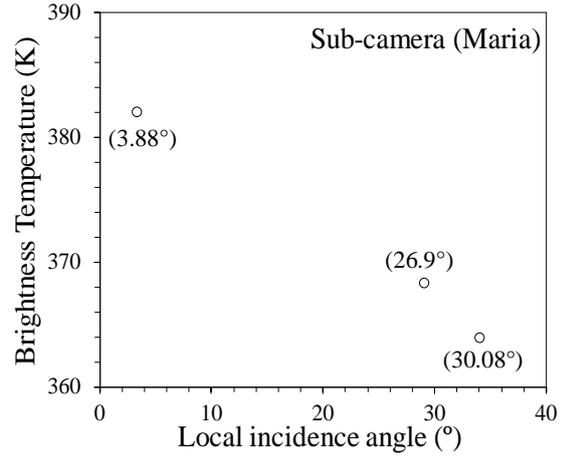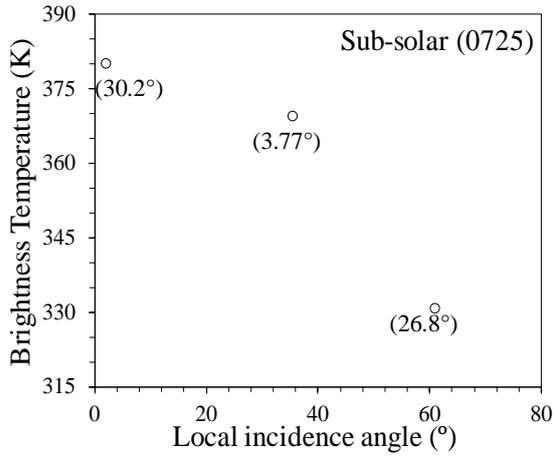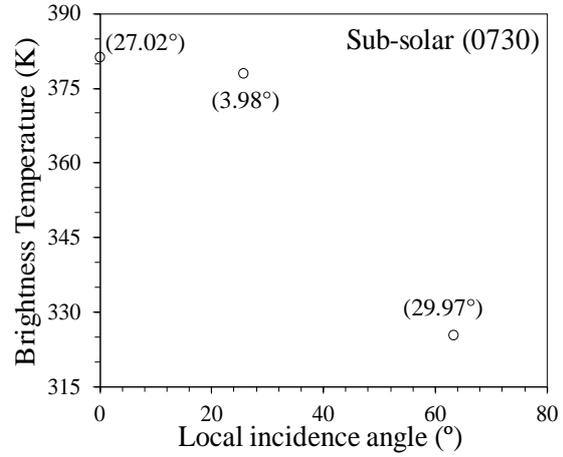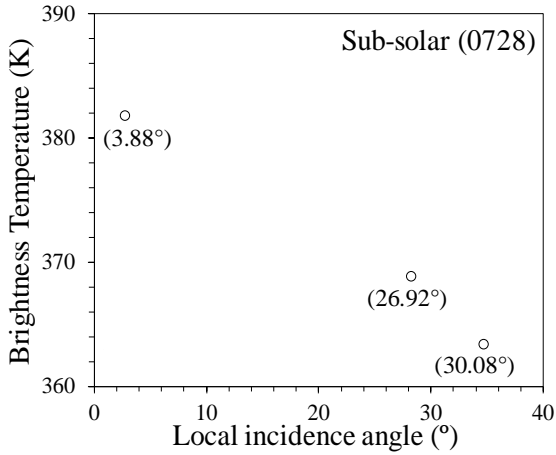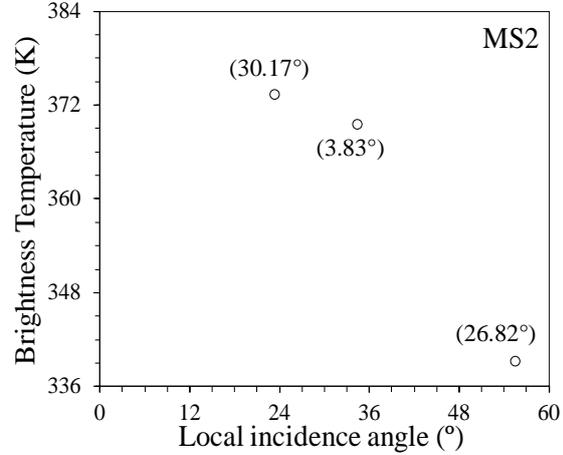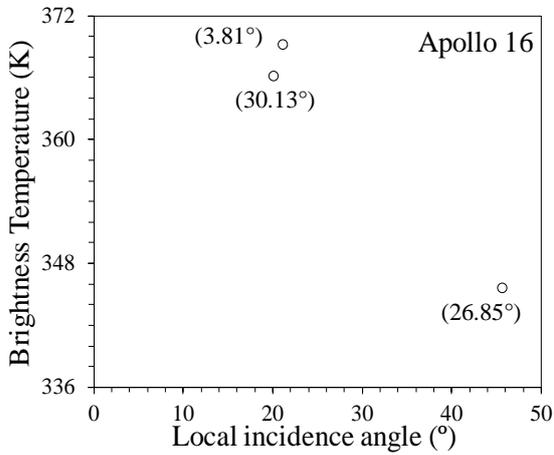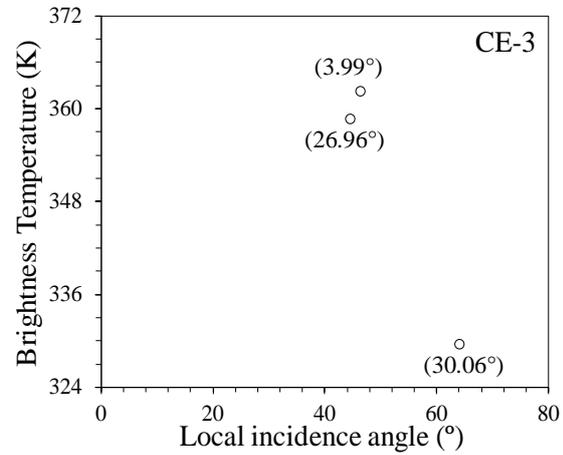

**Figure 6.** Brightness temperature versus local incidence angle for eight typical areas. The two sub-solar points of July 25 and 28 are from nearby homogeneous maria areas to avoid fresh craters which are the exact sub-solar points. Angles in the bracket are phase angles.

**Appendix A**

**The calibration and brightness temperature derivation**

To acquire the brightness temperature map of the Moon using the GF-4 MIR band, we performed 1) an absolute radiometric calibration using the Moon as the source and 2) a separation of the reflected sunlight and thermal emission of the Moon. The data acquired on March 2 and August 4 were not calibrated because their exposure times were different from those acquired in latter days (March 2 was the first occasion acquiring MIR Moon data and was useful for optimizing the exposure time for future observations). Only the images with exposure time of 4 ms for the MIR band and 30 ms for the VNIR bands were calibrated.

The GF-4 VNIR bands were calibrated using Chang'E-1 Interference Imaging Spectrometer (IIM) data because its stability has been demonstrated by its seamless mosaic (Wu et al. 2013; 2018). The pixels of both IIM and GF-4 data with similar composition, illumination and observation geometry (solar incidence angle, observation angle and phase angle) were extracted using a combination of computer code and manual operation. The human check was to ensure that both the area of the box for averaging IIM and GF-4 data were the same and to select a smooth area from a large number of candidates with similar geometry. The same was done for the MIR band. The IIM spectra were resampled to GF-4 bands using the following equation:

$$r_b = \frac{\int_{\lambda_1}^{\lambda_2} r_\lambda S_b(\lambda) d\lambda}{\int_{\lambda_1}^{\lambda_2} S_b(\lambda) d\lambda} \quad (2)$$

where $r_b$ is the resampled spectra of GF-4 band, $r_\lambda$ is the original radiance spectra of IIM, and $S_b(\lambda)$ is the spectral response function of the GF-4 band. The relationship between GF-4 digital numbers (DN) and the resampled radiance was constructed. To ensure a reliable calibration, the calibration was also performed using the Moon Mineralogy Mapper ($M^3$) data by different people with a similar approach. Due to the existence of systematic bias in the absolute reflectance among different $M^3$ optical periods, the $M^3$ data acquired during the Optical Period 2C1 (OP2C1) were used considering that this OP matches other data such as IIM and the in situ spectra measured by the Chang'E-3 rover spectrometer based on large numbers of comparisons (Wu et al. 2018). Figure 7 shows the calibration plot of GF-4 band 5 using the IIM and M³ data. Both calibration lines, which are mutually consistent, are linear with an R² of 0.98 and an uncertainty of ± 4.55% (1σ standard deviation).

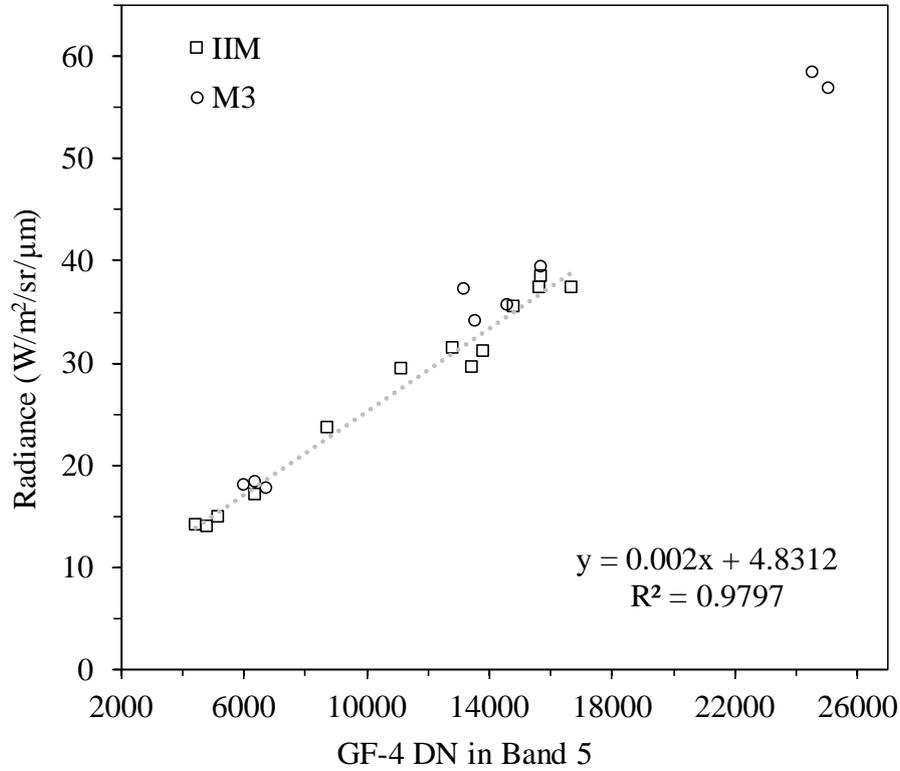

**Figure 7.** On-board calibration plot of GF-4's band 5 using the IIM and M³ data. The function in the plot was derived from IIM data.

The observed radiance of the MIR band consists of a mixture of reflected sunlight and thermally emitted radiance. To calibrate the MIR band the following formula was used:

$$L(\alpha, i, e, \lambda, T) = E_0(\lambda)\text{r}(\alpha, i, e, \lambda) + \varepsilon(e, \lambda)B(\lambda, T) \qquad (3)$$

where $L(\alpha, i, e, \lambda, T)$ is the radiance, $E_0(\lambda)$ is the solar irradiance corrected for the Sun–Moon distance, $\text{r}(\alpha, i, e, \lambda)$ is the bidirectional reflectance, $\varepsilon(e, \lambda)$ is the directional emissivity, and $B(\lambda, T)$ is the thermally emitted radiance of a blackbody at temperature $T$.

The bidirectional reflectance varies strongly with the illumination and observation geometry. Moreover, both the solar incidence angle and observation angle vary considerably across the lunar disk. Here, a novel method which need not consider angles was developed, taking benefit from the simultaneous imaging of all six bands.

The basis of this method is to find a relationship between the reflectance in the MIR band and a VNIR band (e.g., band 5). Hence, the reflected sunlight radiance of the MIR band can be derived from the VNIR band. The directional hemispherical reflectance (DHR) spectra of lunar samples better obey Kirchhoff's laws than the bidirectional reflectance spectra. These spectra are from the Johns Hopkins University Spectral Library and the Relab at Brown University (http://www.planetary.brown.edu/relab/). The DHR spectra were used to investigate the relation between the reflectance in the VNIR and MIR wavelengths. The spectra of the two datasets were resampled to GF-4 bands using Equ. (2). Figure 8 shows that the reflectance relationship is linear between the NIR and MIR bands. Hence the reflected radiance of the MIR band can be derived.

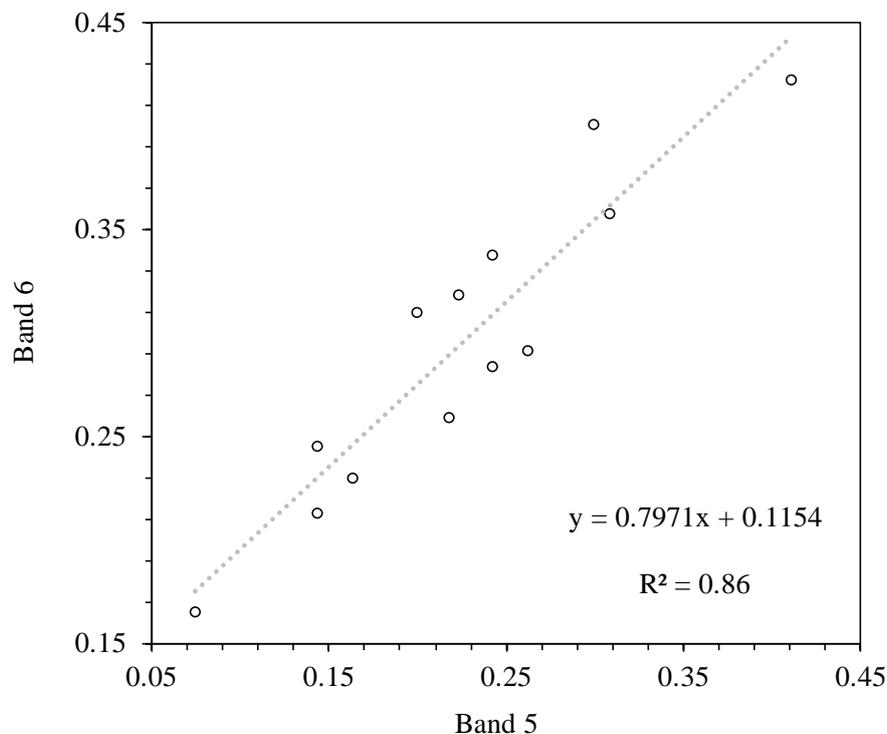

**Figure 8.** Scatter plot of GF-4's band 5 and band 6 reflectance.

The emissivity of the MIR band was also estimated from the Johns Hopkins University Spectral Library. It is known that the reflectance measured in the laboratory is much higher than the reflectance of the actual lunar surface (Pieters et al. 2008). The reflectance of the Johns Hopkins University Spectral Library was corrected using a correction factor of 0.535 which was derived from dividing the laboratory-measured reflectance of a mature lunar soil (Apollo bulk soil 62231) by the remote sensing reflectance of the Apollo 16 calibration site. The emissivity of 0.77 for highlands and

0.82 for maria were derived from Kirchhoff's law and the corrected reflectance.

To model the temperature, a 1- dimensional thermal model was used as follows:

$$\rho(z)C(T)\frac{\partial T(z,t)}{\partial t} = \frac{\partial}{\partial z}\left[k(z,T)\frac{\partial T(z,t)}{\partial z}\right] \qquad (4)$$

$$\varepsilon\sigma T_0^4(t) = F(t) + k(z,T)\frac{\partial T(z,t)}{\partial z}|_{z=0} \qquad (5)$$

$$F(t) = \frac{S}{d^2}\cos(i)(1-A) \qquad (6)$$

in which $S$ is the solar constant, $d$ is the Sun–Moon distance, $A$ is the bolometric albedo, $i$ is the solar incidence angle, $\varepsilon$ is the wavelength-integrate hemispherical emissivity, and $\sigma$ is the Stefan-Boltzmann constant. $F(t)$ represents the solar insolation and the last term of Equ. (5) represents the subsurface conduction. The thermal conductivity $k(z,T)$, capacity $C(T)$, and density $\rho(z)$ of lunar soils are from Bauch et al. (2014) which were originally from Apollo samples. $\varepsilon$ is 0.95 for highlands and 0.96 for maria. The bolometric albedo $A$ varies from 0.06 to 0.2 through the consideration of the location of the calibration points and review of previous results (Bandfield et al. 2015; Racca 1995; Williams et al. 2011; Wohler et al. 2017). Figure 9 shows the calibration plot of GF-4 band 6. The calibration line is linear with $R^2$ of 0.98 and uncertainty ± 2.25% (1σ standard deviation).

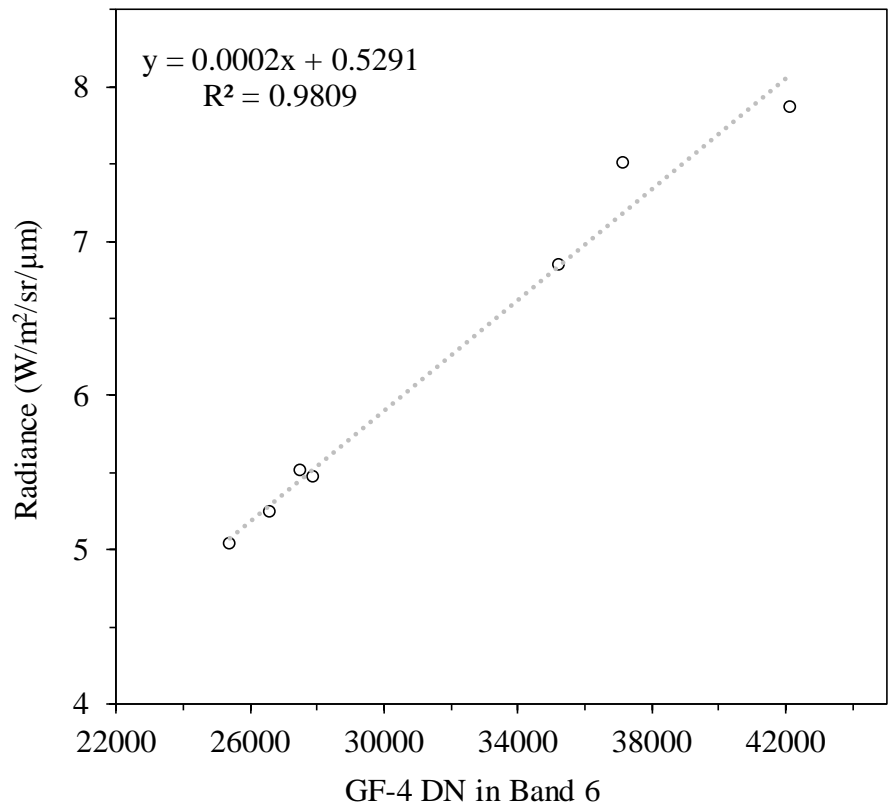

**Figure 9.** On-board calibration plot of GF-4's band 6.

**Appendix B**

**The images of all the 5 days**

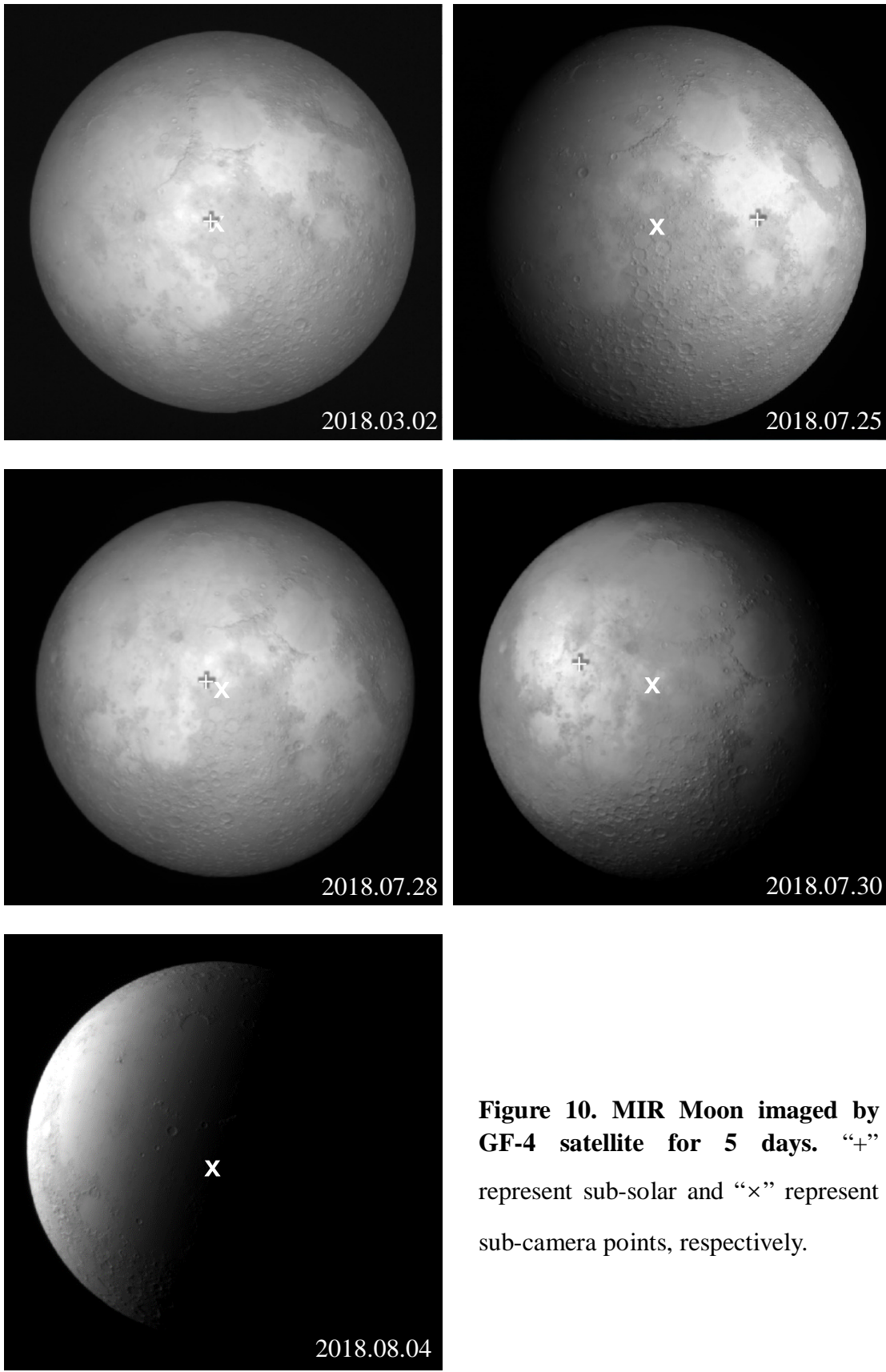

Figure 10. MIR Moon imaged by GF-4 satellite for 5 days. "+" represent sub-solar and "×" represent sub-camera points, respectively.

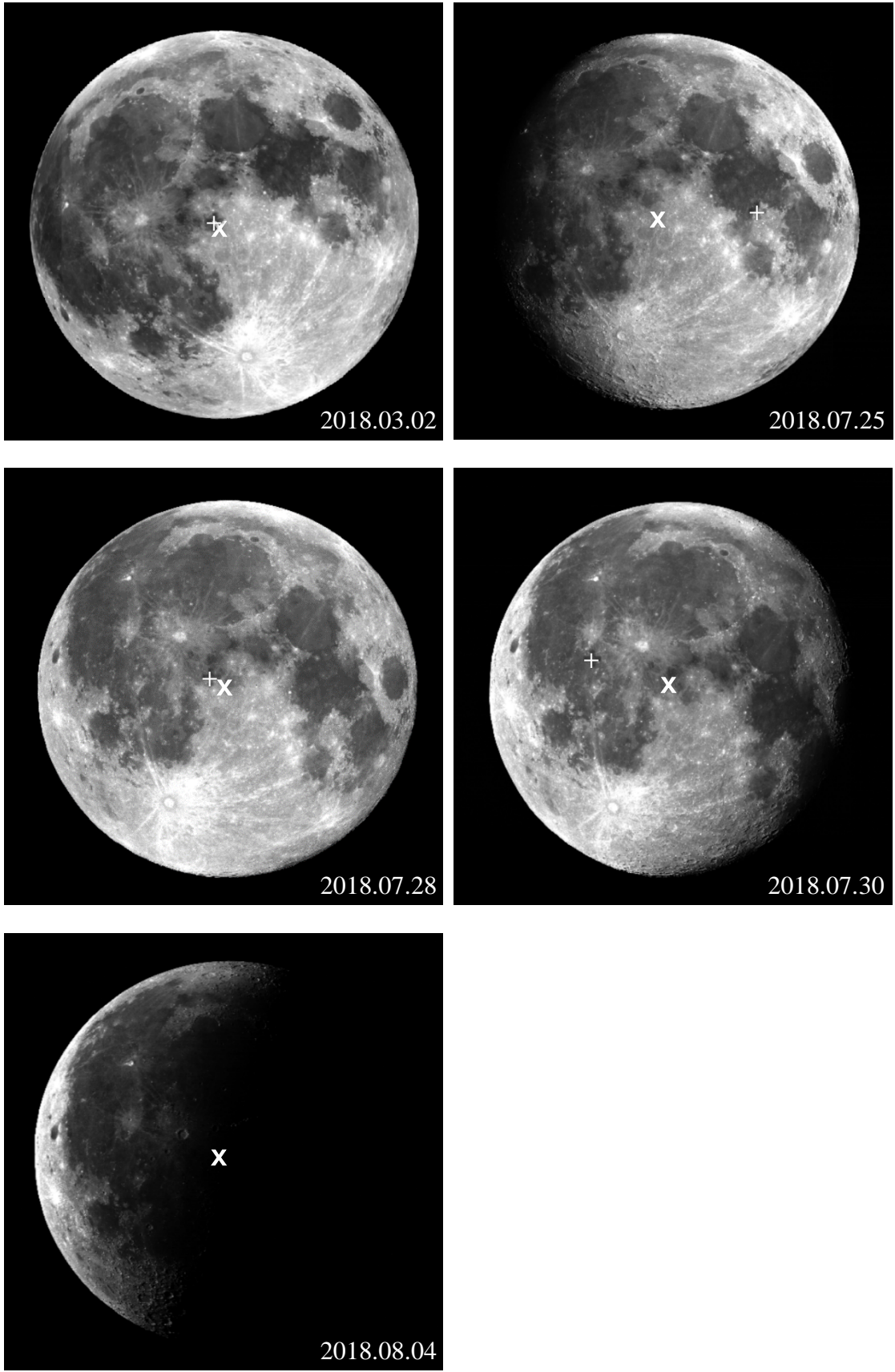

**Figure 11. NIR Moon simultaneously imaged by GF-4 satellite as those MIR Moon.**

"+" represent sub-solar and "×" represent sub-camera points.

**Appendix C**

**Radiance versus local incidence angle for eight typical areas of Band 5**

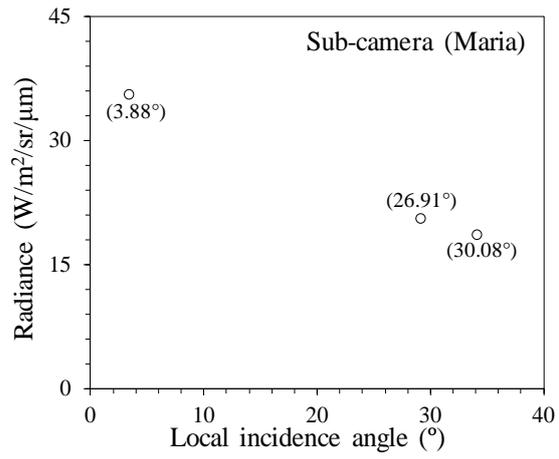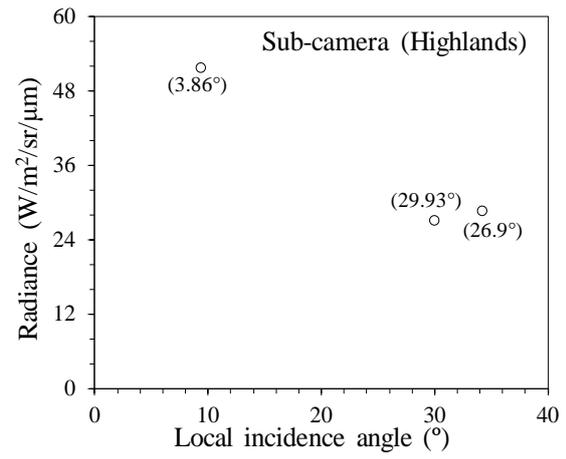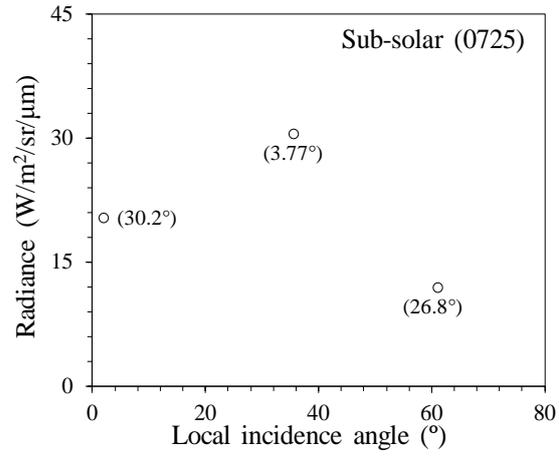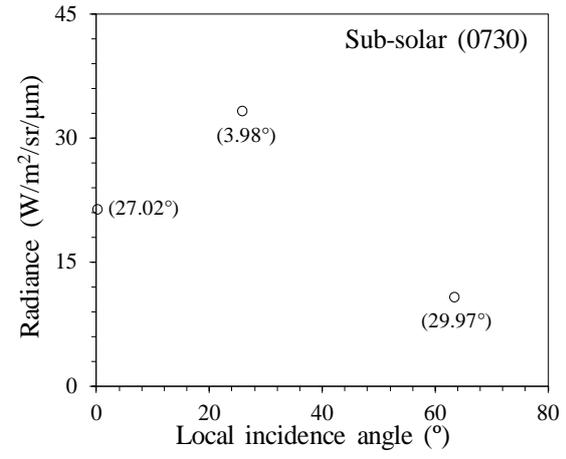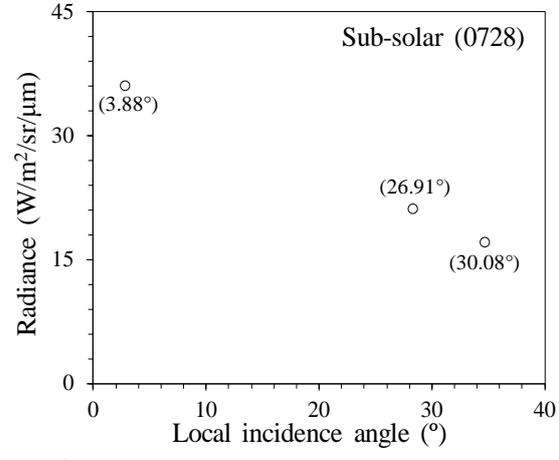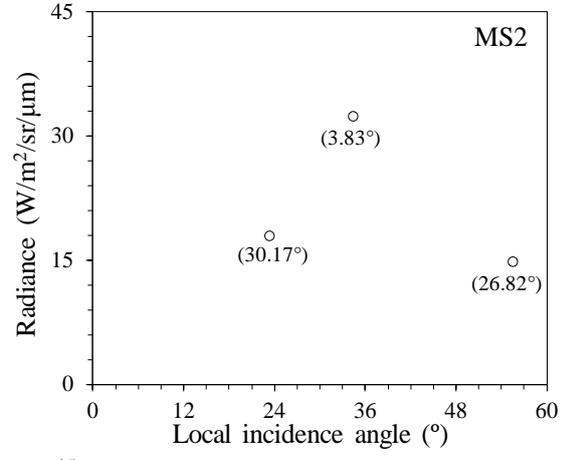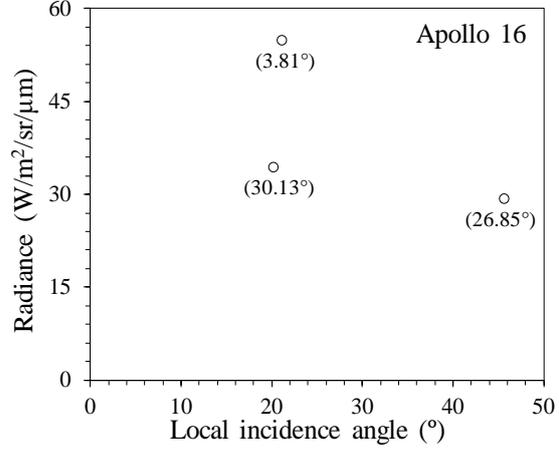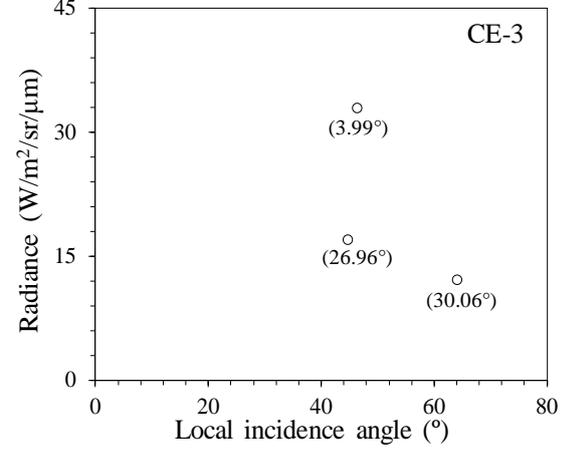

**Figure 12.** Radiance versus local incidence angle for eight typical areas of Band 5. The two sub-solar points of July 25 and 28 are from nearby homogeneous maria areas to avoid fresh craters which are the exact sub-solar points.